\documentclass[useAMS,usenatbib,referee]{biom}

 \newcommand{\bitem}{\begin{itemize}}
 \newcommand{\eitem}{\end{itemize}}

\newcommand{\diag}{  {\rm diag}  }
\newcommand{\logit}{  {\rm logit}  }
\usepackage{multicol} 
\usepackage{color}
\usepackage{amsmath}
\usepackage[pdftex]{graphicx}
\usepackage{amssymb}
\usepackage[font=small,labelfont=bf]{caption} 
\usepackage[figuresright]{rotating}
\usepackage{algorithm}
\usepackage{algpseudocode}
\usepackage[english]{babel} 
\usepackage{tikz}
\usetikzlibrary{shapes,decorations,arrows,calc,arrows.meta,fit,positioning}
\tikzset{
    -Latex,auto,node distance =1 cm and 1 cm,semithick,
    state/.style ={ellipse, draw, minimum width = 0.7 cm},
    point/.style = {circle, draw, inner sep=0.04cm,fill,node contents={}},
    bidirected/.style={Latex-Latex,dashed},
    el/.style = {inner sep=2pt, align=left, sloped}
}

\PassOptionsToPackage{hyphens}{url}
\usepackage[colorlinks = true,
            linkcolor = blue,
            urlcolor  = blue,
            citecolor = blue,
            anchorcolor = blue]{hyperref}

\usepackage[figuresright]{rotating}

\title[Machine Learning Algorithms for Causal Inference in Radiomics Data Analysis]{Ultra-high dimensional confounder selection algorithms comparison with application to radiomics data}

\author
{Ismaïla Bald\'e $^{1, 2, *}$ and Debashis Ghosh $^{2, **}$ 
\email{ismaila.balde@umoncton.ca} \\
\email{debashis.ghosh@cuanschutz.edu} \\
$^1$ Department of Mathematics and Statistics, Université de Moncton, Moncton, NB, E1A 3E9, Canada \\
$^2$  Department of Biostatistics and Informatics, Colorado School of Public Health, Aurora, CO, 80045, USA}

\begin{document}

\pagerange{\pageref{firstpage}--\pageref{lastpage}} 
\volume{}
\pubyear{2023}
\artmonth{October}

\doi{-}

\label{firstpage}
\begin{abstract}
Radiomics is an emerging area of  medical imaging data analysis particularly for cancer. It involves the conversion of digital medical images into mineable ultra-high dimensional data. Machine learning algorithms are widely used in radiomics data analysis to develop powerful decision support model to improve precision in diagnosis, assessment of prognosis and prediction of therapy response. However, machine learning algorithms for causal inference have not been previously employed in radiomics analysis. In this paper, we evaluate the value of machine learning algorithms for causal inference in radiomics. We select three recent competitive variable selection algorithms for causal inference: outcome-adaptive lasso (OAL), generalized outcome-adaptive lasso (GOAL) and causal ball screening (CBS). We used a sure independence screening procedure to propose an extension of  GOAL and OAL for ultra-high dimensional data, SIS + GOAL and SIS + OAL. We compared SIS + GOAL, SIS + OAL and CBS using simulation study and two radiomics datasets in cancer, osteosarcoma and gliosarcoma. The two radiomics studies and the simulation study identified SIS + GOAL as the optimal variable selection algorithm. 
\end{abstract}


\maketitle

\section{Introduction \label{section intro}}
Recent advancements in the field of medical image analysis have resulted in the development of an emerging technology, namely, Radiomics. Radiomics is the process that allows the conversion of digital medical images into mineable data. It extracts ultra-high dimensional sets of imaging features which can be used to build appropriate statistical models to assist in diagnosis, prognosis and therapy monitoring (Qian et al., 2021; Zhang et al., 2021). Radiomics analyses have been shown to yield models that improve precision in diagnosis, assessment of prognosis and prediction of therapy response. A major challenge in radiomics analysis is the curse of dimensionality (Qian et al., 2021). It is well established that  the curse of dimensionality often induces the collinearity problem (Zou and Zhang, 2009). 

In ultra-high dimensional data analysis, variable selection algorithms have been proven to be successful machine learning techniques to select important features while removing irrelevant features to improve statistical efficiency (Zou, 2006; Zou and Zhang, 2009; Qian et al., 2021; Tang et al., 2022). More recently, variable  selection algorithms have been used to improve prediction modeling in radiomics analysis (Van Timmeren et al., 2020; Qian et al., 2021; Zhang et al., 2021; Ghosh et al., 2022; Zhong et al., 2022). However, variable  selection algorithms for causal inference, which select true confounders and precision variables to improve confounding bias and statistical efficiency, have not been previously used in radiomics data analysis. The variable selection problem is often encountered in causal inference from observational data (Robins and Greedland, 1986; Brookhar et al., 2006;  De Luna et al., 2011; Patrick et al., 2011; Wang et al. (2012);  Zigler and Dominici, 2014; Wilson and Reich, 2014; Shortreed and Ertefaie, 2017; Ertefaie et al., 2018; Antonelli et al. 2019; Tang et al., 2022; Baldé et al., 2023). 

In observational studies, such as radiomics data analysis, there often exists a set of baseline covariates (confounders) related to both exposure and outcome of interest. An ideal causal method has to adjust for all true confounders to avoid bias. In a seminal paper, Rosenbaum and Rubin (1983) proposed the propensity score (PS) methodology to remove confounding bias.  The PS is defined as the probability of receiving the treatment given the vector of baseline covariates. While traditional approaches used expert knowledge of the observed data for both outcome and propensity score models specifications, the true structure of the observed data is generally unknown in practice  and therefore must be estimated (Tang et al., 2022).

Variable selection for causal inference is a growing popular topic. In the last two decades, several variable selection algorithms have been proposed to target appropriate variables to construct an unbiased and efficient PS estimator. One of the most popular algorithms in the last ten years is the outcome-adaptive lasso (OAL: Shortreed and Ertefaie, 2017). Baldé et al. (2023) generalized OAL for high dimensional data or even low dimensional data with correlated predictors, GOAL. Tang et al. (2022) proposed the causal ball screening (CBS) to target appropriate covariate for ultra-high dimensional data.

Our  new contributions in the literature are two-fold: 

\begin{enumerate}
 \item We apply variable selection algorithms for causal inference to radiomics data analysis, which have not been  employed in previous work in the literature.
\item We extend the OAL (Shortreed and Ertefaie, 2017) and GOAL (Baldé et al. 2023) algorithms to ultra-high dimensional data by using the sure independent screening proposed by Tang et al. (2022) in the first step. 
\end{enumerate}

The rest of the paper is structured as follows. We present motivating radiomics examples in Section 2. In Section 3, we briefly present the three competitive data-driven algorithms for ultra-high dimensional causal inference, namely SIS + OAL, SIS + GOAL and CBS. We study the statistical theory of the SIS + GOAL and SIS + OAL procedures in Section 4. In Section 5, we describe results of a simulation study comparing methods. In Section 6, we compare SIS + OAL, SIS + GOAL and CBS using real radiomics datasets. We close with some discussion in Section 7.

\section{Radiomics data in cancer\label{section data}}

In this paper, we will use two different radiomics datasets to compare the three variable selection algorithms for ultra-high dimensional causal inference, namely, CBS, SIS + OAL and SIS + GOAL. We describe the osteosarcoma study  in Section 2.1 and the gliosarcoma study Section 2.2.

\subsection{Osteosarcoma study \label{osteo}} 
Osteosarcoma is the most common malignant bone primary cancer. It mostly occurs  in children, adolescents and young adults (Zhang et al., 2021;  Ghosh et al., 2022; Zhong et al., 2022). This cancer usually develops in the osteoclast, the cells that form bone. Recently, many researchers have focused on the diagnosis and the treatment of osteosarcoma. Neoadjuvant chemotherapy (NAC) has improved the $5$-years survival rate from $20-30\%$ to $60-80\%$ (Zhang et al., 2021). More recently,  Zhang et al. (2021) performed a study with 102 patients with osteosarcoma who underwent NAC. Zhang et al. (2021) used the Radcloud software platform to extract $1\, 409$ quantitative imaging features, which can be divided into four groups:

   ($i$) \textbf{group A}: typical summaries for the distribution of voxel intensities within the MR image;
  ($ii$) \textbf{group B}: three dimensional features that reflect the shape and size of the region;
     ($iii$) \textbf{group C}: second order texture features that quantify region heterogeneity differences, calculated from gray-level run length and gray-level co-occurence texture matrices;
    ($iv$) \textbf{group D}: $1\, 302$ first-order statistics and texture features after applying Laplacian, logarithmic, exponential, and wavelet filters on the image.

Previous studies have established an association between surgical stage and treatment response of NAC (Zhang et al., 2021). In this osteosarcoma data, we study whether surgical stage affects the treatment response of NAC, adjusting for  radiomics features as confounders. 

\subsection{Gliosarcoma study\label{glio}} 
Glioblastoma or glioblastoma multiforme (GBM) is the most common and aggressive primary brain cancer (Miller, 2007; Tamimi and Juweid, 2017). GBM usually develops in the glial cells (Ohgaki, 2009; Ghosh et al.,  2022). Gliosarcoma (GSM) is one of the distinct morphological variants of GBM recognized by the  World Health Organization  (Miller, 2007; Qian et al, 2021; Mirchia et al, 2023). GSM  is a rare cancer of  the central nervous system (Mirchia et al, 2023). In a recent paper, Qian et al. (2021) conducted a study with a sample size $n=183$ patients including 100 with GBM ($58$ males and $42$ females) and $83$ with GSM ($58$ males and $25$ females) with $1\, 303$ radiomic features extracted from MRI images. The age range for GSM patients was $16-77$ years and for GBM patients was $12-77$. We took the variable ``Edema" (Yes: $1$ / No: $0$) as the exposure variable (treatment) and response variable $Y$ ($Y = 1$ if the subject had gliosarcoma and $Y = 0$ if the subject had glioblastoma). 
\section{ Methods\label{method}}
In the sequel, we define a binary treatment (exposure) $A$, a continuous outcome $Y$ and baseline covariate $\mathbf{X}=(X_1, \ldots,X_p)$. We focus on estimating the average causal effect $ATE=E[Y(1)-Y(0)]$, where $Y(1)$ and $Y(0)$ denote the potential outcome under treatment and control, respectively. We assume that the design matrix $\mathbf{X}$ ultra-high dimensional in the sense of Tang et al. (2022).

Causal inference from observational studies is based on several assumptions. The following four assumptions are required to guarantee unbiased estimators of the ATE: positivity, consistency, exchangeability and stable unit treatment value assumption. 
  $\mathcal{A}_1$: Positivity can be written as $0<P(A=1|\mathbf{X}=x)<1$ for all possible $x$ values. It means that the probability of receiving both levels of treatment conditional on $\mathbf{X}$ is positive for all individuals. $\mathcal{A}_2$: Consistency is defined as $Y=AY(1)+(1-A)Y(0)$. That is, the observed outcome for an individual is equal to the counterfactual outcome under the treatment assignment the individual actually received. $\mathcal{A}_3 $: Exchangeability or no unmeasured confounding assumptions $\mathbf{X}$ includes all possible confounders: $\{Y(1), Y(0)\} \amalg A \mid \mathbf{X}$. $\mathcal{A}_4$: Stable unit treatment value assumption means each individual's counterfactual outcomes are not influenced by the treatment status of other individuals: $\{Y_i(1), Y_i(0)\} \amalg A_s$, for $i \neq s$.
\subsection{Conditional ball covariance screening [Tang et al., 2022]\label{ball}}
We briefly recall the sure independence screening procedure of Tang et al., (2022) which they called conditional ball covariance screening .
Let $\delta_{ij,k}^X=I\{ X_k \in \bar{B}_{\rho}(X_i,X_j) \}$, where $I(.)$ is an indicator function. Let $\delta_{ij,kl}^X=\delta_{ij,k}^X \delta_{ij,l}^X$ and $\delta_{ij,klst}^X=(\delta_{ij,kl}^X+\delta_{ij,st}^X-\delta_{ij,ks}^X-\delta_{ij,lt}^X)/2$. One can similarly define $\delta_{ij,k}^Y, \delta_{ij,kl}^Y$ and $\delta_{ij,klst}^Y$.
Tang et al., (2022) defined the empirical conditional ball covariance $Bcov_n(\mathbf{X}, Y | A)$ as the square root of: 
$$Bcov^2_n(\mathbf{X}, Y | A)=\frac{\hat{w}}{n_1^6}\sum_{(i,j,k,l,s,t): A_i,A_j,A_k,A_l,A_s,A_t=1} \xi^X_{ij,klst} \xi^Y_{ij,klst} $$
$$+ \frac{1-\hat{w}}{n_0^6} \sum_{(i,j,k,l,s,t): A_i,A_j,A_k,A_l,A_s,A_t=1} \xi^X_{ij,klst} \xi^Y_{ij,klst},$$
where $n_1$ is the number of patients treated, $n_0$ is the number of patients untreated and $\hat{w}=n_1/n$ is the empirical estimator of $w$ which is the probability of receiving treatment. The full SIS procedure is available in  the original paper (Tang et al., 2022: Section 3). The SIS is also discussed in Fan and Lv (2008), Pan et al. (2018), Pan et al. (2020), Barut et al. (2016) and Zou and Zhang (2009).
\subsection{Ultra-high dimensional confounder selection algorithms\label{algo}}
We present the extension of OAL (Shortreed and Ertefaie, 2017) and GOAL (Baldé et al., 2023) for ultra-high dimensional data. Specifically, we combine the sure independence screening proposed by Tang et al. (2022), and OAL or GOAL. We call these procedures SIS + OAL and SIS + GOAL, respectively. GOAL refers to the GOALi version in Baldé et al (2023), which was recommended when the dimensions is high.  
We assume the propensity score model defined as 
\begin{equation}
 \logit\left\lbrace \pi ( \mathbf{X},\alpha)\right\rbrace=\logit\left\lbrace P(A=1\vert \mathbf{X})\right\rbrace= \sum_{j=1}^p \alpha_jX_j. 
 \label{psori}
\end{equation}
Let $\mathcal{C}$ denote the indices of confounders, which are defined as variables that are both associated to outcome and treatment. Let $\mathcal{P}$ denote the indices of precision variables, which are defined as variables that are only associated to the outcome. Let $\mathcal{I}$ denote the variables that are only associated to the treatment. Let $\mathcal{S}$ denote the indices of variables that are unrelated to both outcome and treatment.
Define $\mathcal{A}= \mathcal{C} \cup \mathcal{P}$ and $\mathcal{A}^c= \mathcal{I} \cup \mathcal{S}$. 
The goal of our proposed procedures is to estimate the following PS model 
$$\logit\left\lbrace \pi ( X,\alpha)\right\rbrace=\sum_{j\in \mathcal{A}} \alpha_jX_j. $$
In each procedure, we first apply SIS to reduce the dimension to $q<n$. Then, we fit the data $(\mathbf{X}_\mathcal{K},A)$ by using OAL or GOAL; where $\mathcal{K}=\{ j=1,\ldots,q\}$. The negative log-likelihood is given by $\ell_n (\alpha; A,\mathbf{X}_\mathcal{K})=\sum_{i=1}^n \left\lbrace-a_i(x_{i, \mathcal{K}}^T \alpha_{\mathcal{K}})+\log\left( 1+e^{x_{i, \mathcal{K}}^T \alpha_{\mathcal{K}}} \right) \right\rbrace$.
 The SIS + OAL estimator is defined as:
  $$\hat{\alpha}_\mathcal{K}(SIS + OAL)= \arg \min_{\alpha_\mathcal{K}}  \left[ \ell_n (\alpha; A,\mathbf{X}_\mathcal{K})+ \lambda_1 \sum_{j=1}^q \hat{w}_j|\alpha_{j,\mathcal{K}} | \right],$$
 and  the SIS + GOAL estimates are:
 $$\tilde{\alpha}_\mathcal{K}(SIS + GOAL)= \arg \min_{\alpha_\mathcal{K}}  \left[ \ell_n (\alpha; A,\mathbf{X}_\mathcal{K})   + \lambda_1 \sum_{j=1}^q \hat{w}_j|\alpha_{j,\mathcal{K}} | + \lambda_2 \sum_{j=1}^q \alpha_{j,\mathcal{K}}^2 \right],$$
where 
$\hat{w}_j=\left|{\hat{\beta}_j}\right| ^{-\gamma}$ with  $\gamma>1$, for $j=1,\ldots,q$
and
$(\hat{\beta}_A,\hat{\beta})=\arg \min_{(\beta_A,\beta)}  \mathcal{L}_n^Y \left(\beta_A,\beta; Y, A,\mathbf{X}_\mathcal{K}\right)$;
 $\mathcal{L}_n^Y$ is the negative log-likelihood of the outcome $Y$ given the treatment $A$ and the design matrix $\mathbf{X}_\mathcal{K}$ parametrized by $(\beta_A,\beta)$ for a sample of size $n$. $\hat{\beta}_A$ is the coefficient estimate corresponding to the treatment and $\hat{\beta}$ are the coefficient estimates corresponding to all the $q$ covariates.
The proposed algorithm that extends GOAL for ultra-high dimensional data is as follows. 
\newpage
\newcounter{algsubstate}
\renewcommand{\thealgsubstate}{\alph{algsubstate}}
\newenvironment{algsubstates}
  {\setcounter{algsubstate}{0}%
   \renewcommand{\State}{%
     \stepcounter{algsubstate}%
     \Statex {\footnotesize\thealgsubstate:}\space}}
  {}
\renewcommand{\thealgorithm}{}
{\tiny{
\begin{algorithm}[H]
  \caption{\textbf{SIS + GOAL (GOAL: Baldé et al., 2023)} }\label{goalpirls}
  \begin{algorithmic}[1]
     
\State \textbf{INPUT}: Given original data $(\mathbf{X},A,Y)$;
     \vspace{0.1cm}
\State \textbf{SIS} (refer to the conditional ball covariance screening of Tang et al., 2022)
      \begin{algsubstates}
      \vspace{0.1cm}
      
\State For $j=1,\ldots,p$, compute $\hat{\rho}_j=BCov^2_n(X_j, Y | A)$;
\vspace{0.1cm}

\State Select the $q$ variables with the largest $\hat{\rho}_j$, and denote them as $\mathcal{K}$; without loss of generality, let $\mathcal{K}=\{ j=1,\ldots,q\}$;
       \end{algsubstates}
        \vspace{0.1cm}

\State \textbf{GOAL} (Baldé et al., 2023)
      \begin{algsubstates}
          
     \State For each fixed $\lambda_2$ define:  $\mathbf{X}^*_\mathcal{K}= \begin{pmatrix}
\mathbf{X}_\mathcal{K} \\
\sqrt{\lambda_2 } \mathbf{I}_q
\end{pmatrix}$ and $
A^*=\begin{pmatrix}
A \\
0_q \\
\end{pmatrix};
$
\vspace{0.1cm}

\State Initialize $\tilde{\alpha}_\mathcal{K}$ to $0$;
\vspace{0.1cm}

\State Compute $\tilde{p}(x_{i, \mathcal{K}})=\frac{1}{1+\exp(-x_{i, \mathcal{K}}^T\tilde{\alpha}_\mathcal{K})}$, $t_i=\tilde{p}(x_{i, \mathcal{K}})[1-\tilde{p}(x_{i, \mathcal{K}})]$, $z_i=x_{i, \mathcal{K}}^T\tilde{\alpha}_\mathcal{K}+\frac{a_i-\tilde{p}(x_{i, \mathcal{K}})}{\tilde{p}(x_{i, \mathcal{K}})(1-\tilde{p}(x_{i, \mathcal{K}}))}$, $i=1,2,\ldots, n$;
\vspace{0.1cm}

\State Set $Z^*=\begin{pmatrix}
Z \\
0_q \\
\end{pmatrix}
$ and $\mathbf{T^*}=\begin{pmatrix}
\mathbf{T} & 0_{n \times q} \\
0_{n \times q}^T &  \mathbf{I_q}
\end{pmatrix}$, 
where $Z=(z_1, \ldots, z_n)^T$, $\mathbf{T}=\diag(t_1,\ldots,t_n)$;

\vspace{0.1cm} 
\State   Call \textbf{OAL algorithm} with augmented data $(\mathbf{X}^*_\mathcal{K},A^*)$ to solve  $$ \tilde{\alpha}_{I, \mathcal{K}}^* \mbox{(naive adaptive elastic net)} = \arg \min_{\alpha_\mathcal{K}}  \left[ \ell_{Q^*}(\alpha_\mathcal{K}; \mathbf{X}^*_\mathcal{K},A^*,Z^*,\mathbf{T^*})  + \lambda_1 \sum_{j=1}^q \tilde{w}_j|\alpha_{j, \mathcal{K}} | \right];$$  
\State Compute $\tilde{\alpha}_{I, \mathcal{K}}\mbox{(adaptive elastic net)}=(1+\lambda_2)\tilde{\alpha}_{I, \mathcal{K}}^*\mbox{(naive adaptive elastic net)}$;
\vspace{0.2cm}

\State Update $\tilde{\alpha}_\mathcal{K}=\tilde{\alpha}_{I, \mathcal{K}} \mbox{(adaptive elastic net)}$; 
\vspace{0.1cm}

\State Repeat c $-$ g until convergence of  $\tilde{\alpha}_\mathcal{K}$;
\vspace{0.1cm}

\State Set $\tilde{\alpha}_{I, \mathcal{K}}\mbox{(adaptive elastic net)}=\tilde{\alpha}_\mathcal{K}$;
 \end{algsubstates}
\State \textbf{OUTPUT}: $\tilde{\alpha}_{I, \mathcal{K}}\mbox{(adaptive elastic net)}$.

\end{algorithmic}
\end{algorithm} 
}}
\vspace{-1cm}
Note that $\ell_{Q^*}$ is the quadratic approximation of $\ell_n$ in the augmented data $(\mathbf{X}^*_\mathcal{K},A^*)$. 
Due to space constraints,  the SIS + OAL and CBS (Tang et al., 2022) algorithms are  presented in the Web Appendix A and B, respectively.
\subsection{Tuning parameters selection\label{tuning}}
In this section, we describe the tuning parameters selection which is an important issue in practice. To find the optimal $(\lambda_1, \lambda_2, \gamma)$, we follow the tuning parameters selection procedure of the original algorithms OAL, GOAL and CBS. As seen in several papers (Shortreed and Ertefaie, 2017; Tang et al.,  2022; Baldé et al., 2023), the selected pair of parameters $(\lambda_1, \gamma)$ minimize the weighted absolute mean difference (wAMD). The wAMD (Shortreed and Ertefaie, 2017) is defined as: 
 $$wAMD(\lambda_1;\mathbf{X},A)= \sum_{j=1}^q \left|{\beta_j}\right| \left|   \frac{\sum_{i=1}^n \hat{\tau}_i^{\lambda_1} X_{ij}A_i}{  \sum_{i=1}^n \hat{\tau}_i^{\lambda_1} A_i}  - \frac{\sum_{i=1}^n \hat{\tau}_i^{\lambda_1} X_{ij}(1-A_i)}{ \sum_{i=1}^n\hat{\tau}_i^{\lambda_1} (1-A_i)} \right|.$$
To facilitate the comparison, we use  the same possible values of $\lambda_1$ ($S_{\lambda_1}$) for all algorithms SIS+OAL, SIS+GOAL and CBS. We follow Shortreed and Ertefaie (2017) to set $S_{\lambda_1}$:  $S_{\lambda_1}=\{  n^{-10}, n^{-5},n^{-2}, n^{-1}, n^{-0.75}, n^{-0.5}, n^{-0.25}, n^{0.25}, n^{0.49}\}.$ 
 To select the optimal tuning parameter $\lambda_2$, we follow Baldé et al. (2023) who used a two dimensional wAMD procedure to select the optimal pair $(\lambda_1, \lambda_2)$. For a given $\lambda_2$,  we used the wAMD function to  select the optimal $\lambda_1$. The pair of tuning parameters $(\lambda_1, \lambda_2)$ that minimize the wAMD function is selected for GOAL. Note that $S_{\lambda_2}$ (Baldé et al., 2023) is the possible values of $\lambda_2$ : $S_{\lambda_2}=\{0,10^{-2},10^{-1.5},10^{-1},10^{-0.75},10^{-0.5},10^{-0.25},10^0,10^{0.25},10^{0.5},10^1\}$. We refer the interested readers to the original articles for more details on the tuning parameters selection (Shortreed and Ertefaie, 2017; Tang et al.,  2022; Baldé et al., 2023).

 \section{Statistical theory \label{stheory}} 
 
 In this section, we study the statistical theory of the proposed SIS + OAL and SIS + GOAL procedures. 
 Our procedures are inspired by previous results of Fan and Lv (2008) with SIS + SCAD, Zou and Zhang (2009) with SIS + AEnet and Tang et al. (2022) with CBS (SIS + doubly robust), which have some desirable theoretical guarantees including the oracle property. We now show that our SIS + GOAL and SIS + OAL procedures have the same statistical properties as SIS + SCAD, SIS + AEnet and CBS. 
 Without loss of generality, assume that $\mathcal{A}=\{1, 2, \ldots, q_0\}$ with $q_0<q$. We then write the Fisher information matrix as $$\mathbf{F}(\alpha^*)=
\begin{pmatrix}
\mathbf{F}_{11} & \mathbf{F}_{12}\\
\mathbf{F}_{21} & \mathbf{F}_{22}
\end{pmatrix},$$
where $\mathbf{F}_{11}$ is the Fisher information matrix  (of size $q_0\times q_0$) for the parsimonious PS model.
To develop the statistical theory, some commonly used assumptions and regularity conditions are required, which we give in the Web Appendix C.

The next theorem guarantees that the SIS method may select the $q$ most important variables.

\begin{theorem} \label{SIS}(Sure independence screening, Tang et al., 2022)

Under Conditions (H1)-(H3), we have the following:
\begin{enumerate}
    \item [(a)]
 $P\left( \max_{j\in \mathcal{W}} \hat{\rho}_j< \min_{j\in \mathcal{A}} \hat{\rho}_j\right) \rightarrow 1$, where $\mathcal{W}$ is defined in (H3) (see Web Appendix C) and 
 \item [(b)] $P\lbrace{\mathcal{A}=\left(\mathbf{X}_{\mathcal{C}} \cup \mathbf{X}_{\mathcal{P}}\right) \subset \mathcal{K}\rbrace} \rightarrow 1 \quad \mbox{as} \quad n \rightarrow \infty. $ 
\end{enumerate}
 \end{theorem} 
 Tang et al. (2022) show that all true confounders and pure predictors of the outcome are included with high probability in the set $\mathcal{K}$ selected by the SIS.

We now present our asymptotic results in the following theorems. 

 \begin{theorem} 
 
Under Conditions (H4)-(H8), the SIS + OAL produces an estimator $\hat{\alpha}(SIS + OAL)$ that holds the oracle property; that is $\hat{\alpha}(SIS + OAL)$ must satisfy the following:
\begin{enumerate}
    \item [(i)] Consistency in variable selection: $\lim_n P\{ \hat{\alpha}_j(SIS + OAL)=0|j\in \mathcal{I} \cup \mathcal{S}\}=1$.
    \item [(ii)] Asymptotic normality: $\sqrt{n} \{ \hat{\alpha}(SIS + OAL)-\alpha^*_{\mathcal{A}} \} \rightarrow_d N(0,\mathbf{F}_{11}^{-1})$.
    \end{enumerate}
    \label{proofsisoal}
\end{theorem}
Theorem~\ref{proofsisoal} is a direct application of Theorem 1 in Shortreed and Ertefaie (2017); thus the proof is omitted. It is clear from Theorem 2 that with a proper choice of $\lambda_1$, the SIS + OAL enjoys the oracle property.

\newpage
\begin{theorem} 

 Under Conditions (H4)-(H9), the SIS + GOAL produces an estimator $\hat{\alpha}(SIS + GOAL)$ that holds the oracle property; that is $\hat{\alpha}(SIS + GOAL)$ must satisfy the following:
\begin{enumerate}
    \item [(i)] Consistency in variable selection: $\lim_n P\left\lbrace\hat{\alpha}_j(SIS + GOAL)=0|j\in \mathcal{I} \cup \mathcal{S}\right\rbrace=1$;
    \item [(ii)] Asymptotic normality: $\sqrt{n} \{ \hat{\alpha}(SIS + GOAL)-\alpha^*_{\mathcal{A}} \} \rightarrow_d N(0,\mathbf{F}_{11}^{-1})$.
\end{enumerate}
\label{proofsisgoal}
\end{theorem}
Theorem~\ref{proofsisgoal} is parallel to Theorem 1 in Baldé (2023). Thus, its proof is omitted. By Theorem 3, the SIS + GOAL enjoys the oracle property if the pair of tuning parameters ($\lambda_1$, $\lambda_2$) is chosen appropriately.

\section{Simulation study \label{sim}}
Simulation studies were conducted to compare the performance of the three recent variable selection algorithms for causal inference, namely, SIS + OAL, SIS + GOAL and CBS. We compared the algorithms based on variable selection and estimation accuracy. For variable selection, we evaluated methods based on the probability of each predictor being included in the PS model for estimating the ATE. For estimation accuracy, we compared algorithms based on the bias, standard error (SE) and mean squared error (MSE).
\subsection{Simulation design\label{design}}
In this subsection, we describe the simulation set up to generate the data $(\mathbf{X},A,Y)$, denoting the covariates matrix $(\mathbf{X}=(X_{1}, X_{2},\ldots, 
X_{p}))$, exposure and outcome, respectively. The vector ${X_i} = (X_{i1}, X_{i2},\ldots, 
X_{ip})$, for $i=1, 2, \ldots, n$ is simulated from a multivariate standard Gaussian distribution with pairwise correlation $\rho$. The binary treatment $A$ is simulated from a Bernoulli distribution with $\logit 
\{P(A_i=1)\}=\sum_{j=1}^p \alpha_jX_{ij}$. Given $\mathbf{X}$ and $A$, the continuous outcome is simulated as $Y_i=\beta_A 
A_i+\sum_{j=1}^p \beta_jX_{ij}+ \epsilon_i$ where  $\epsilon_i  \sim N(0,1)$. The true ATE was  
$\beta_A=0$. To examine the performance of different methods, we considered the four $(n,p)$ combinations studied in Tang et al. (2022): $(300,100)$, $(300,1000)$, $(600,200)$ and $(600,2000)$.

In this simulation, we considered the same four scenarios as in Shortreed and Ertefaie (2017), which were also used in Baldé et al. (2023). Let $\beta \in \mathbb{R}^p$ be the regression coefficients in the outcome model and $\alpha \in \mathbb{R}^p$ be the regression coefficients in the exposure model. For each scenario, covariates $X_1$ and $X_2$ are confounders, covariates $X_3$ and $X_4$ are outcome pure predictors, covariates $X_5$ and $X_6$ are exposure pure predictors and the rest ($p-6$) are spurious covariates. We considered two different correlations: independent covariates ($\rho=0$) and highly correlated covariates ($\rho=0.75$). 
The four scenarios are defined as follows (Shortreed and Ertefaie, 2017): 

$\bullet$ Scenario 1: $\beta= (0.6,0.6,0.6,0.6, 0, 0,0, \ldots,0)$ and $\alpha= (1,1,0,0, 1, 1,0, \ldots,0)$;

$\bullet$ Scenario 2: $\beta= (0.6,0.6,0.6,0.6, 0, 0,0, \ldots,0)$ and $\alpha= (\textbf{0.4,0.4},0,0, 1, 1,0, \ldots,0)$;

$\bullet$ Scenario 3:  $\beta= (\textbf{0.2,0.2},0.6,0.6, 0, 0,0, \ldots,0)$ and $\alpha= (1,1,0,0, 1, 1,0, \ldots,0)$;

$\bullet$ Scenario 4: $\beta= (0.6,0.6,0.6,0.6, 0, 0,0, \ldots,0)$ and $\alpha= (1,1,0,0, \textbf{1.8, 1.8},0, \ldots,0)$.

To estimate the ATE, SIS + OAL and SIS + GOAL used the IPTW estimator and CBS used the doubly robust estimator. For variable selection, we conducted 1000 simulations and computed  the proportion of times each variable was selected  for inclusion in the PS model. Each variable was considered selected when the estimated regression coefficient in the PS model using either SIS + OAL, SIS + GOAL or CBS, was greater than the tolerance $10^{-8}$ (Shortred and Ertefaie, 2017). 
We used the bias, standard error (SE) and mean squared error (MSE) to compare SIS + OAL, SIS + GOAL and CBS. We refer readers to Shortreed and Ertefaie (2017) for more details on the simulation and OAL algorithm, to Baldé et al. (2023) for GOAL algorithm and to Tang et al. (2022) for CBS algorithm.

\subsection{Simulation results\label{results}}

\begin{figure}[!h]			
			\begin{center}
			
\vspace{0cm}		
           
  \hspace{0.75cm}     \bf   $\rho=0 $ \hspace{6cm}     $ \rho=0.75$

----------------------------------------------------------------------------------------------

	\includegraphics[width=0.95\linewidth,height=1.2\linewidth]{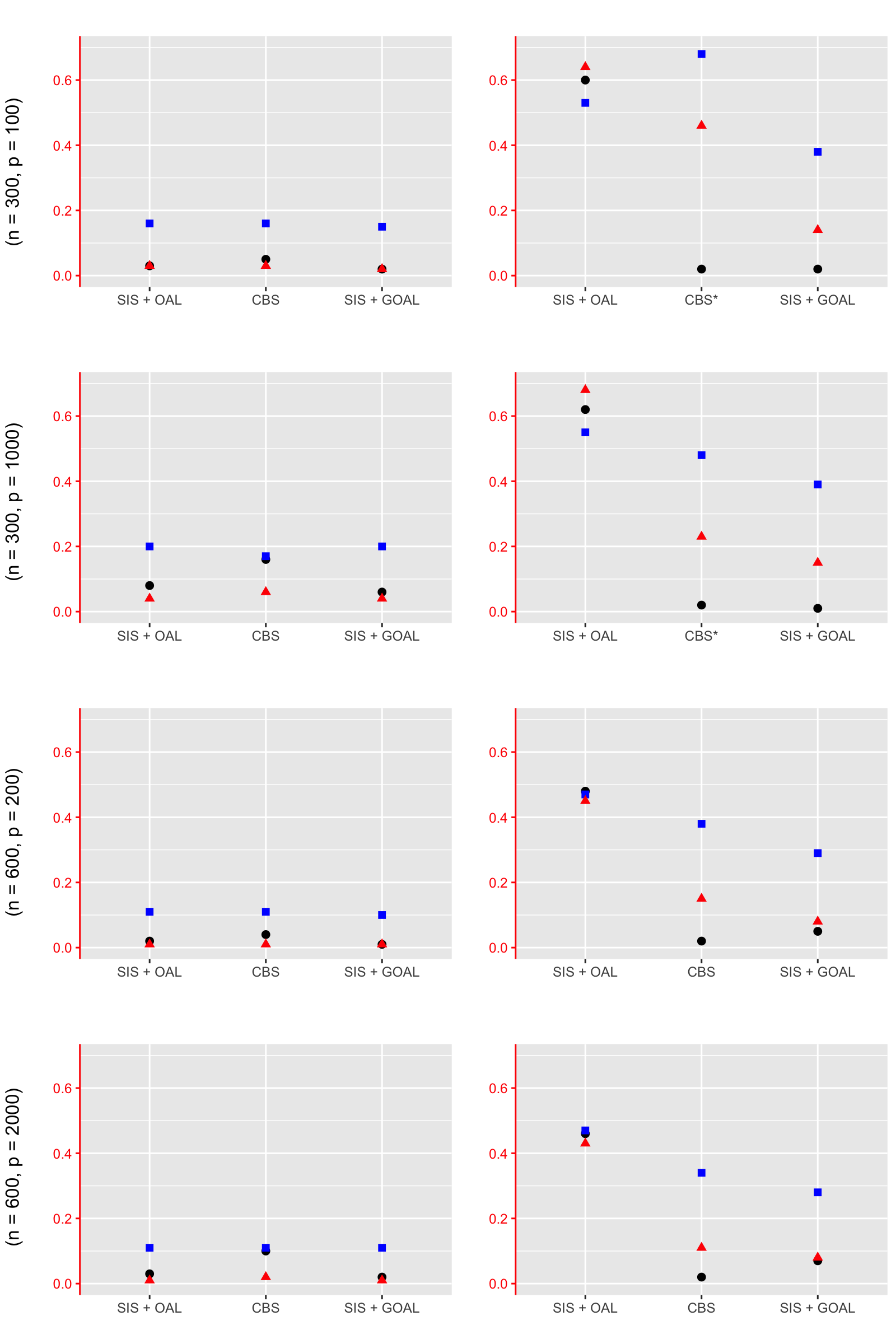}

\caption[]{Absolute bias (circle), standard error (square) and mean squared error (triangle) of IPTW estimator for SIS + OAL and SIS + GOAL and DR estimator for CBS for the  average treatment effect (ATE),  under Scenarios 1 (based on 1000 estimates) for all  $(n,p)$ combinations $(300,100)$, $(300,1000)$, $(600,200)$ and $(600,2000)$ by row. This figure appears in color in the electronic version of this article, and any mention of color refers to that version.\\\\
  \footnotesize{ \textit{Note: CBS had some convergence issues for some $(n, p)$ combinations when $\rho=0.75$. For CBS*, we exclude 9 and 1 simulations for (300, 100)  and (300, 1000), respectively.}}}
\label{simres1}
		\end{center}
  
		\end{figure}

Figure \ref{simres1} displays the bias, standard error (SE) and mean squared error (MSE) of SIS + OAL,  SIS + GOAL and CBS  estimators for the ATE with Scenario 1 in the $(n,p)$ combinations $(300, 100)$,  $(300, 1000)$, $(600, 200)$ and $(600, 2000)$ for both independent $(\rho=0)$ and highly correlated covariates $(\rho=0.75)$.  Due to space constraints in the main manuscript, the corresponding results for Scenarios 2-4 are presented in Web Appendix Section F (refer to Web Figures 3-5). 

In all $(n,p)$ combinations, all three estimators (SIS + OAL,  SIS + GOAL and CBS) performed equivalently when covariates are independent ($\rho=0$).  However, SIS + GOAL performed much better than SIS + OAL and CBS, when covariates are highly ($\rho=0.75$). We found similar results in Scenarios 2-4 (see Web Figures 3-5).

\begin{figure}[H]			
\begin{center}

  \hspace{0.75cm}     \bf   $\rho=0 $ \hspace{6cm}     $ \rho=0.75$
----------------------------------------------------------------------------------------------
 
1		 	\includegraphics[width=7cm,height=5cm]{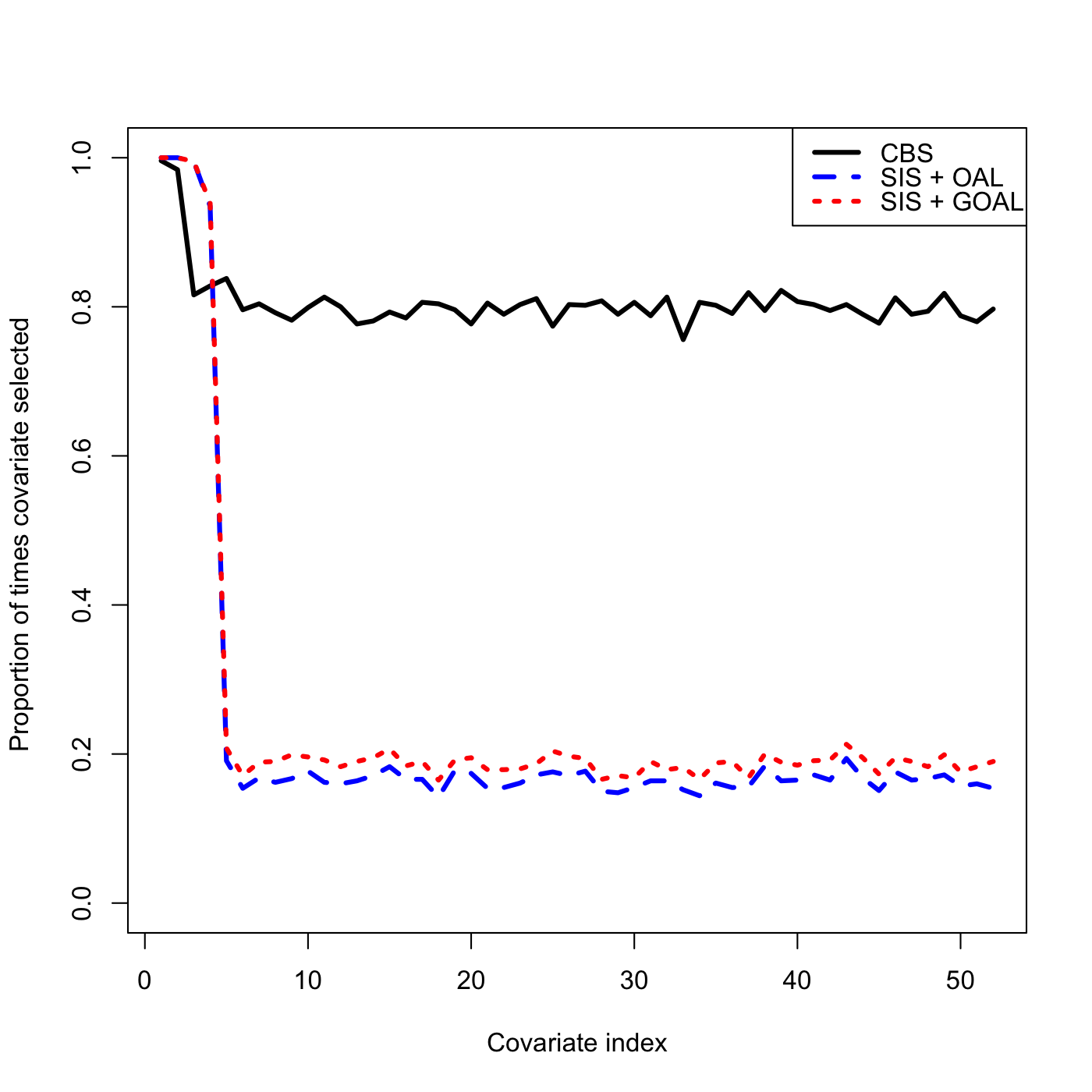}
\hspace{1cm}
			\includegraphics[width=7cm,height=5cm]{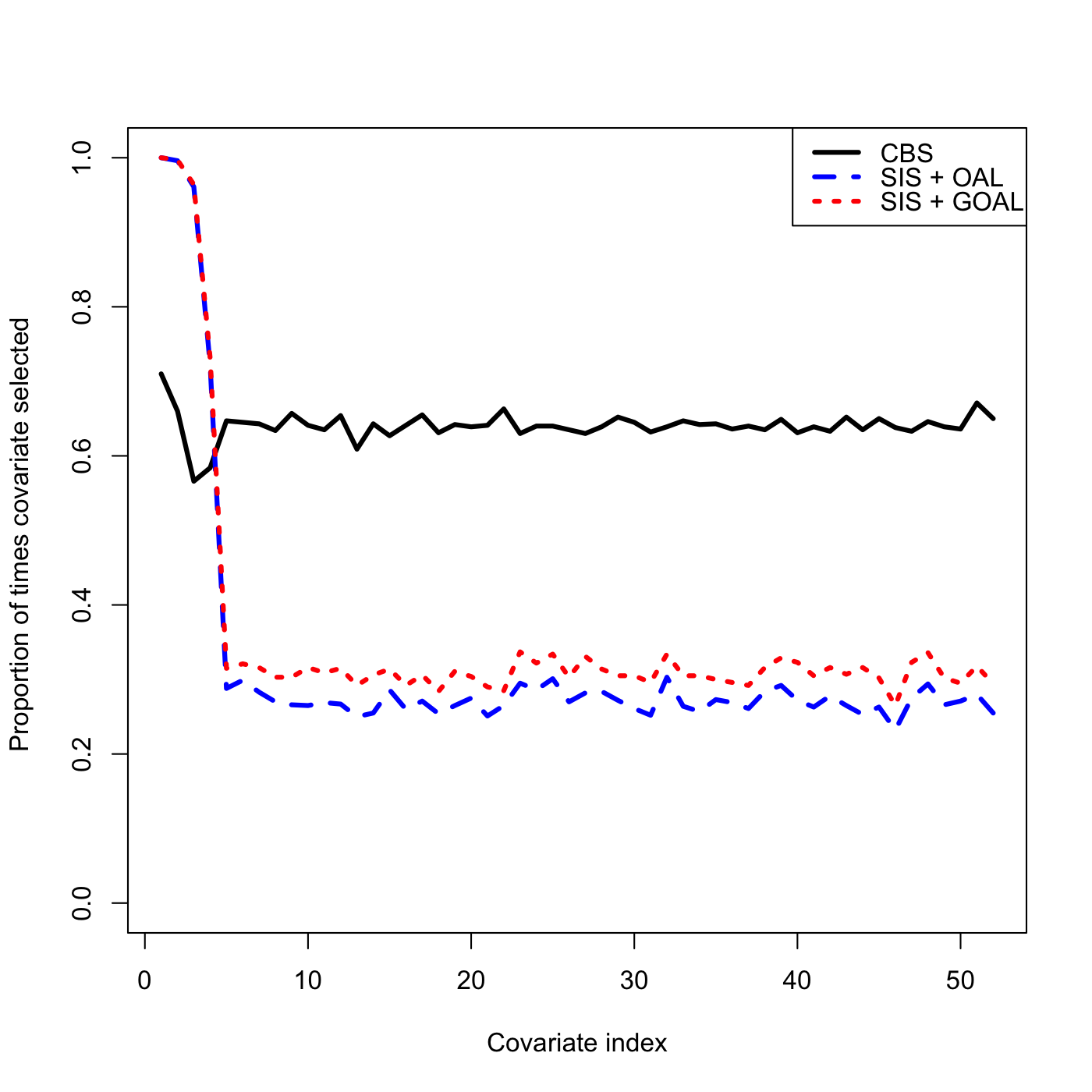}
			               
2			 \includegraphics[width=7cm,height=5cm]{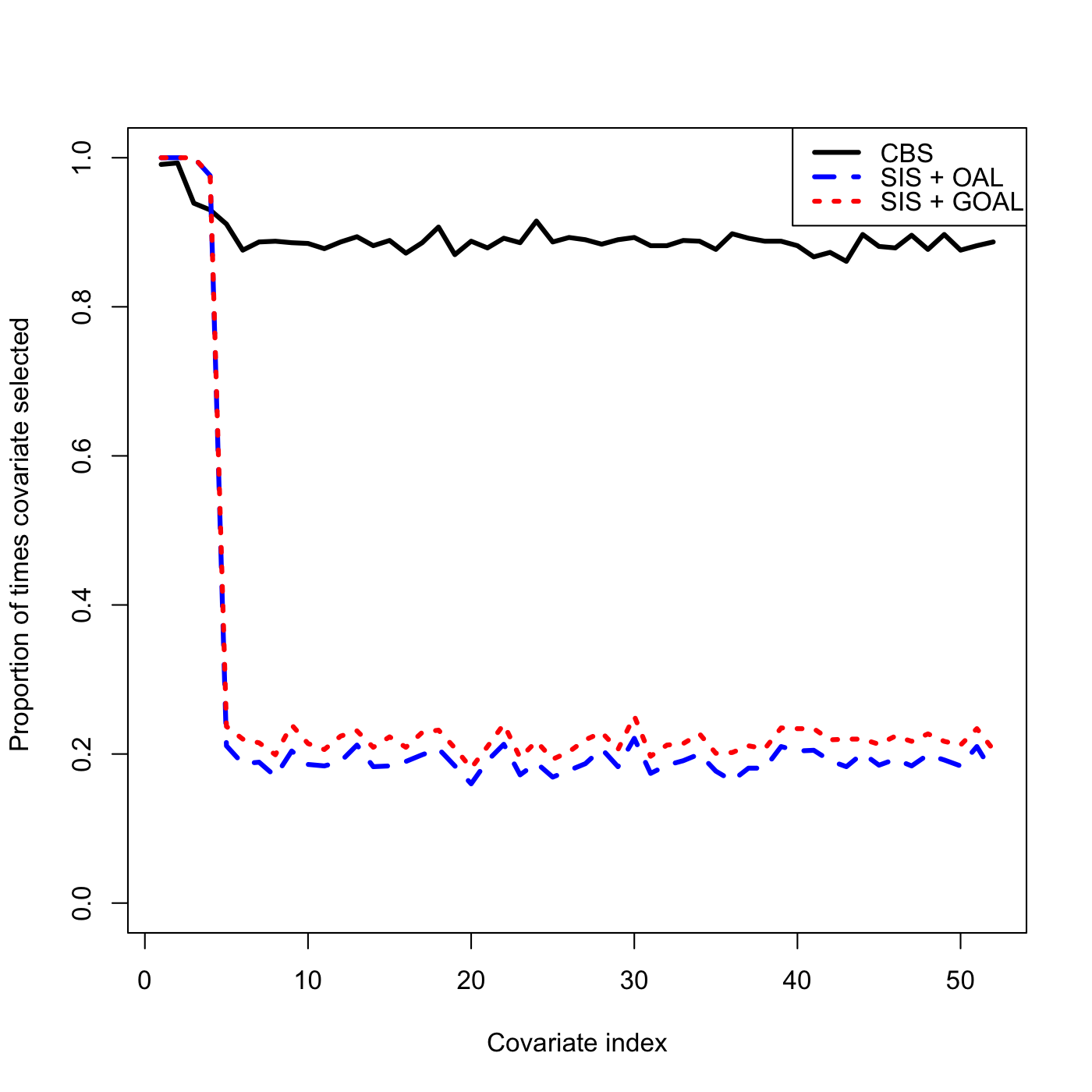}
\hspace{1cm}
			\includegraphics[width=7cm,height=5cm]{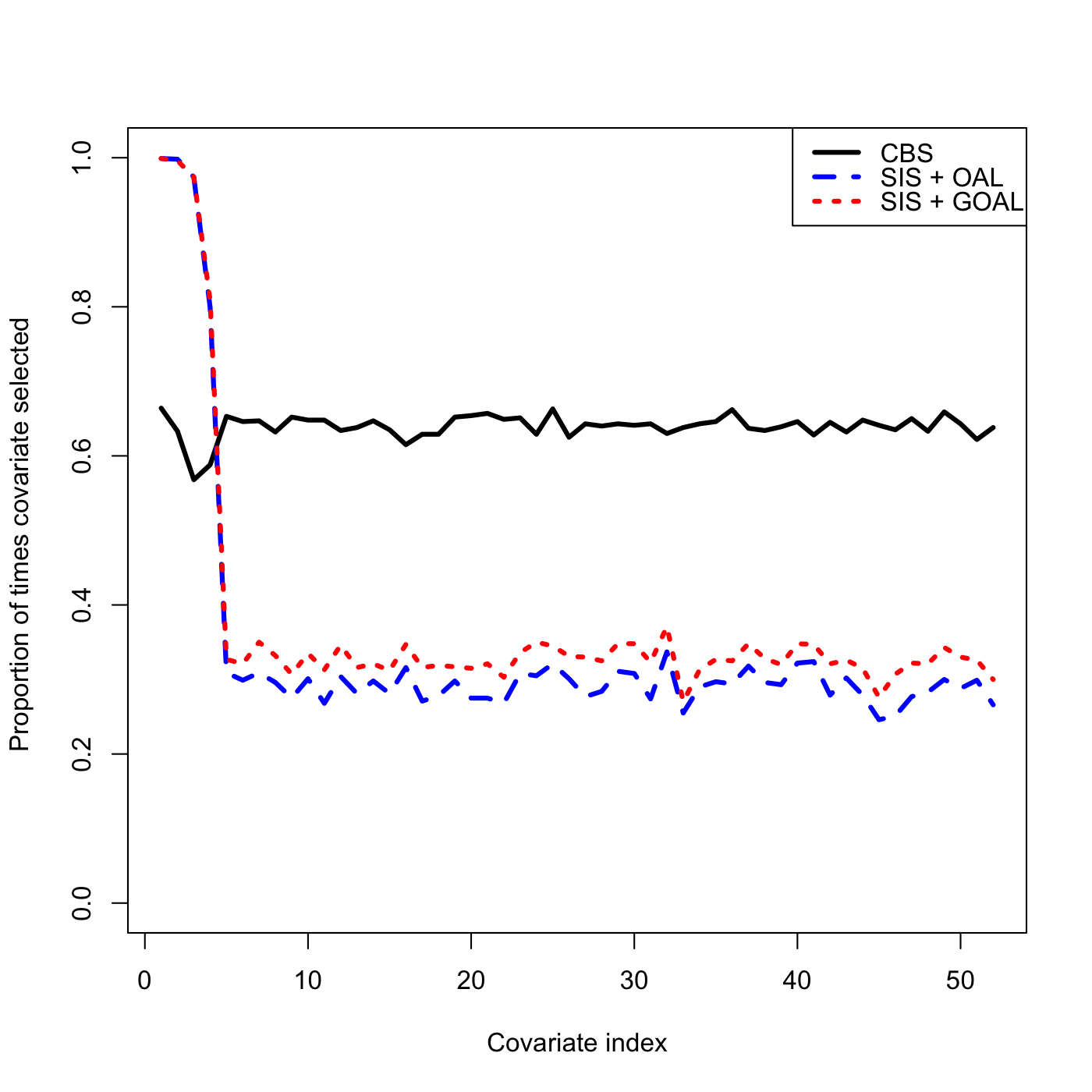}
			
3		 	\includegraphics[width=7cm,height=5cm]{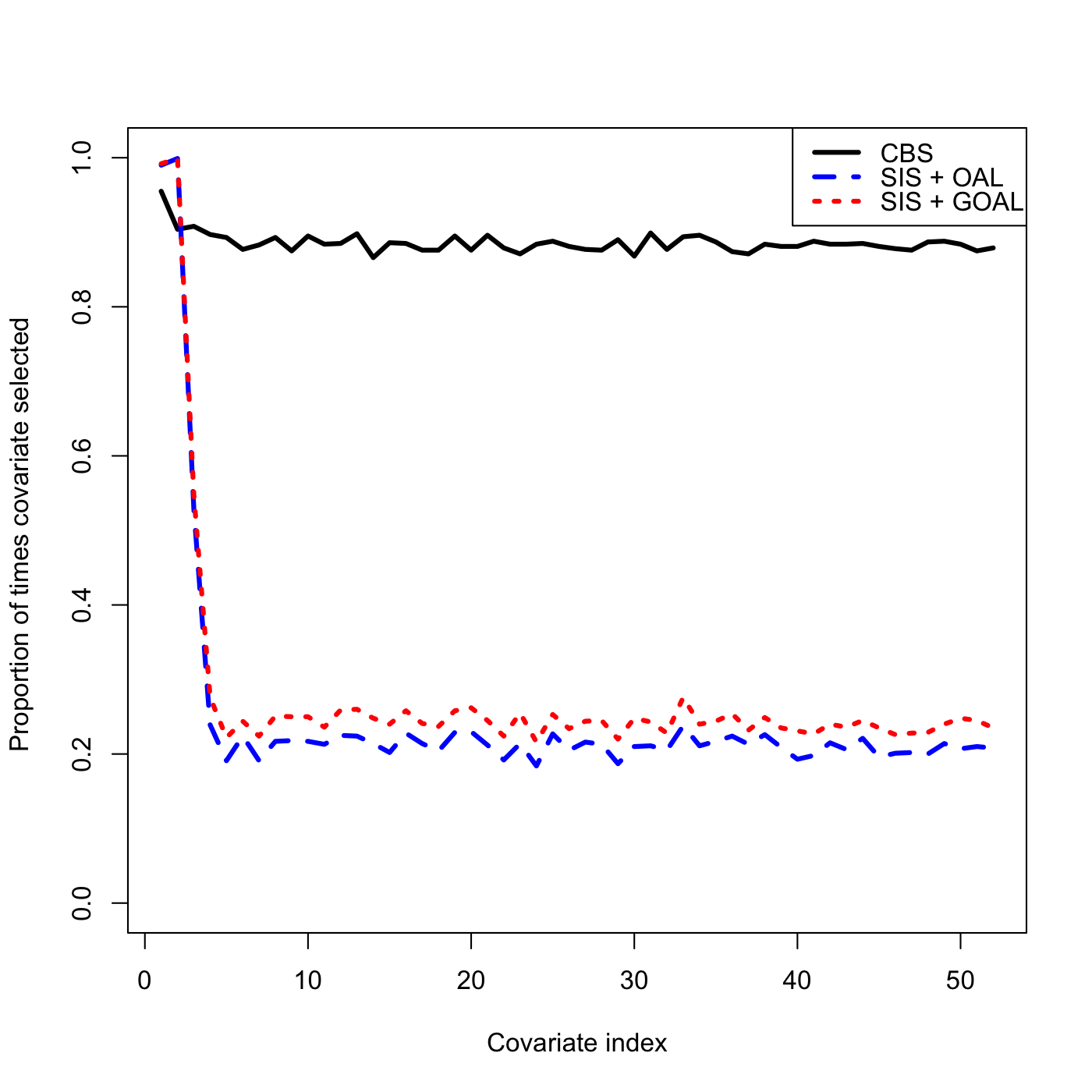}
\hspace{1cm}
			\includegraphics[width=7cm,height=5cm]{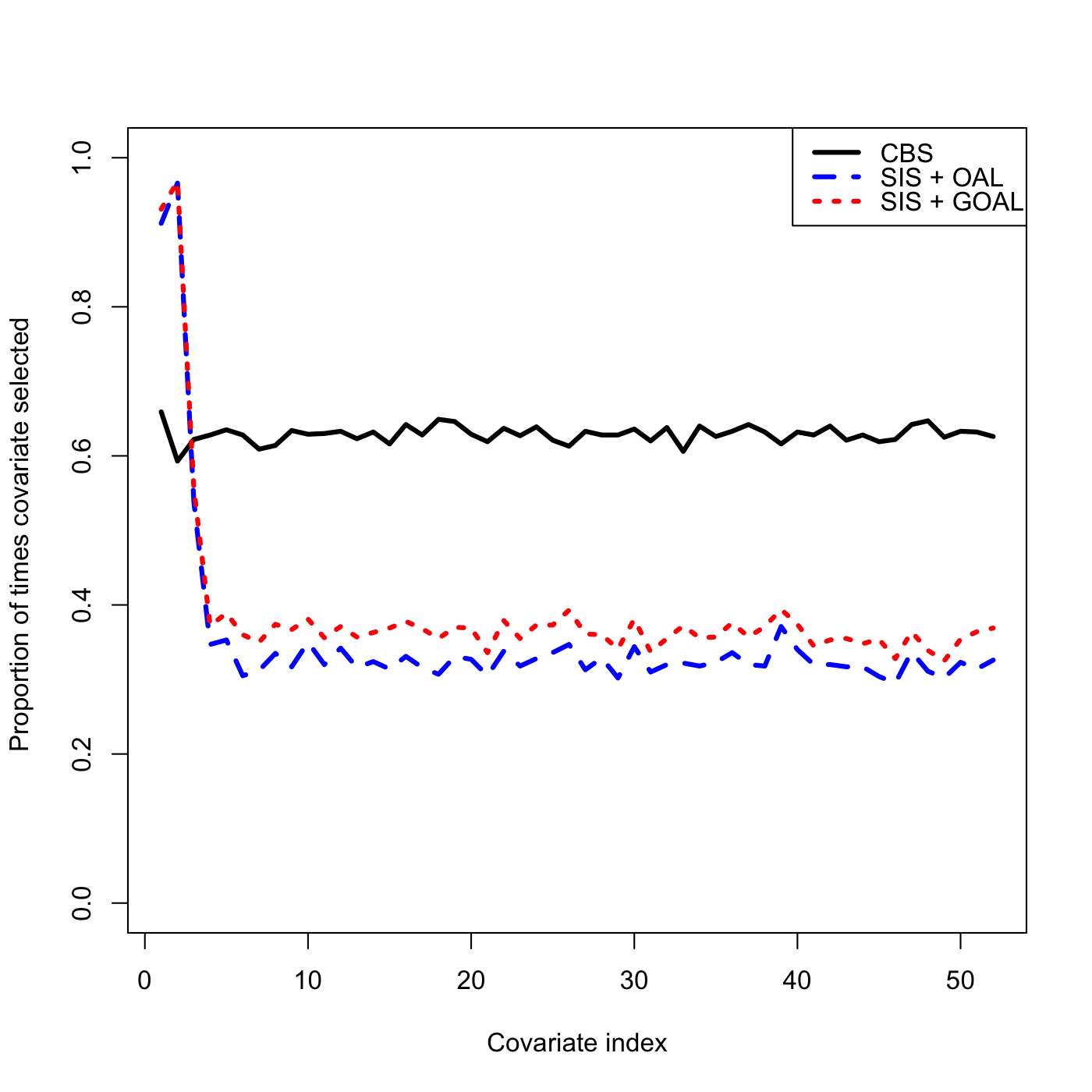}
	
4			 \includegraphics[width=7cm,height=5cm]{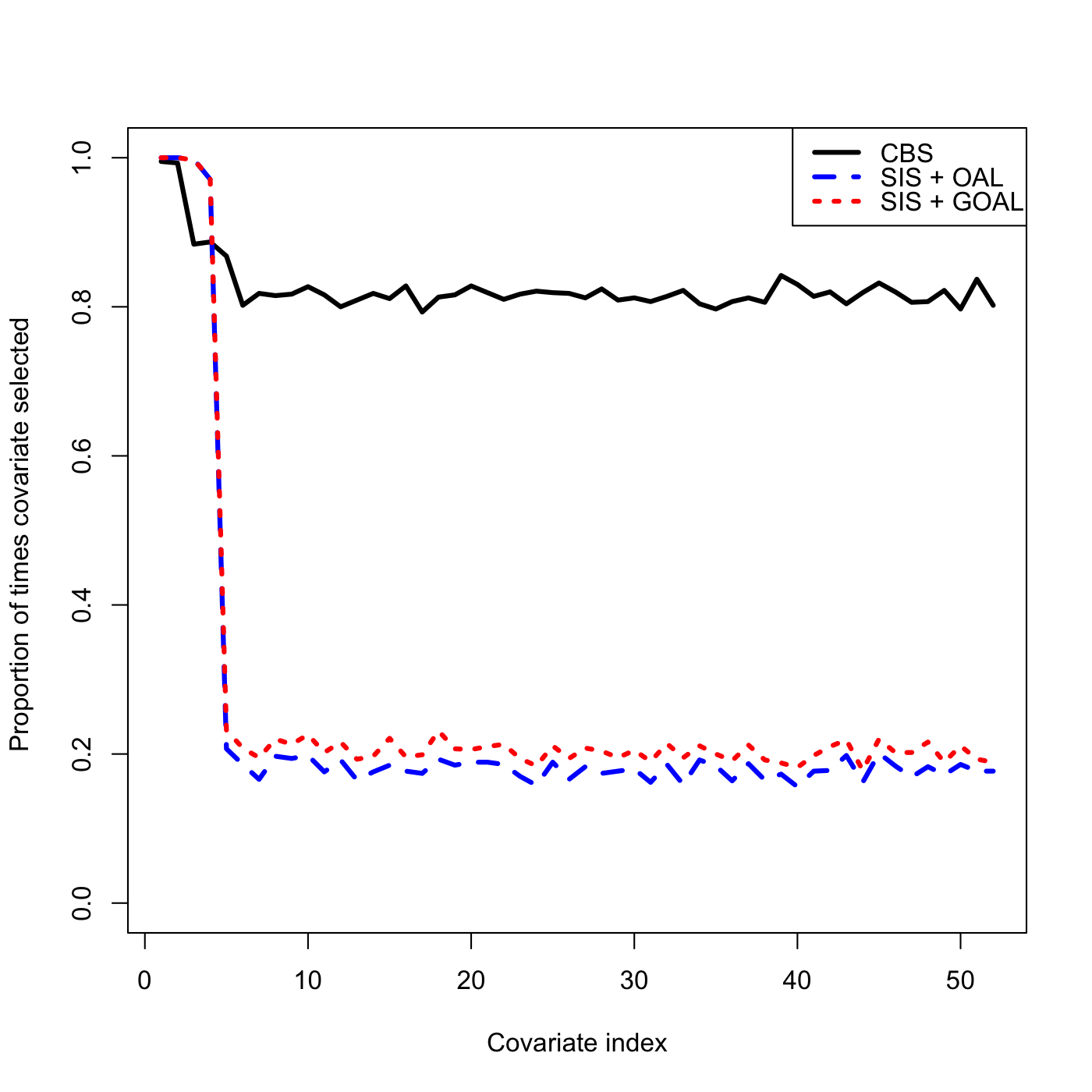}
\hspace{1cm}
			\includegraphics[width=7cm,height=5cm]{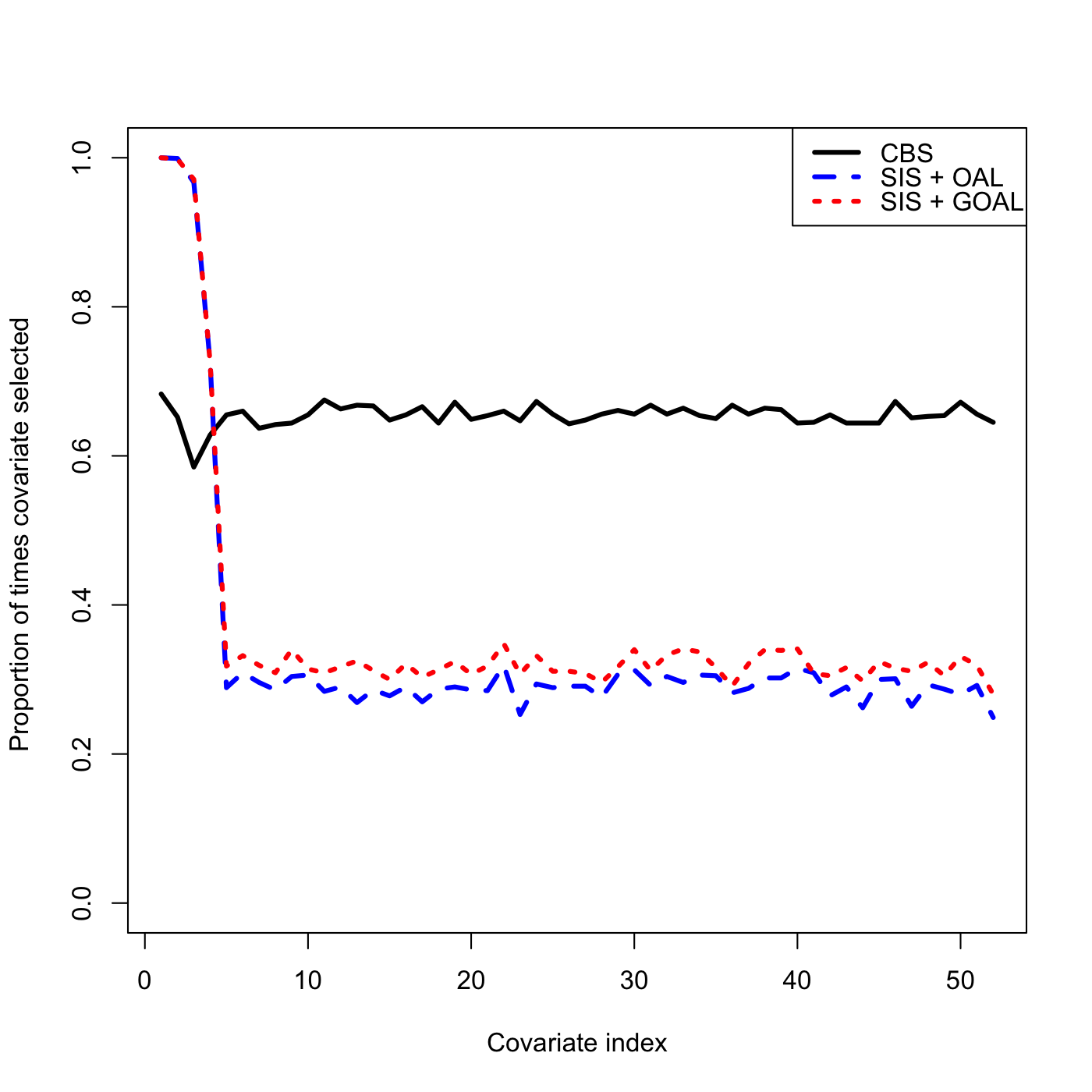}
   \caption[]{Probability of covariate being included in PS model for estimating the average treatment effect (ATE) under Scenarios 1-4 (by row) with sample size $n=300$ and number of covariares $p=1000$. SIS selected $d_n=52$ covariates where covariate index is: 1-2, $C_{AY}$; 3-4, $P_{Y}$; 5-6, $P_{A}$; 7-52, spurious. This figure appears in color in the electronic version of this article, and any mention of color refers to that version.}
\label{simres2}		
		\end{center}
		\end{figure}

Figure \ref{simres2} presents the proportion of times each covariate was selected over $1000$ simulations for inclusion in the PS model (tolerance$=10^{-8}$) of SIS + OAL, SIS + GOAL and CBS for the combination $(n=300, p=1000)$ with both independent $(\rho=0)$ and highly correlated $(\rho=0.75)$ covariates. In all scenarios, SIS + OAL and SIS + GOAL algorithms included all covariates at similar rate with high probability for confounders and pure predictors of the outcome and relatively small probability for the pure predictor of treatment and spurious covariates (between $15$ and $25 \%$ when $\rho=0$ and $25-35\%$ when $\rho=0.75$). However, CBS  included confounders and pure predictors of the outcome less than SIS + OAL and SIS + GOAL  and included the pure predictor of treatment and spurious covariates at least twice more than SIS + OAL and SIS + GOAL. For variable selection, the algorithms SIS + OAL and SIS + GOAL performed well while CBS performed worse.

\section{Real data application: Radiomics data \label{real data}}

\subsection{Osteosarcoma study\label{osteoresults}}
In this example, the exposure is surgical stage (stage III versus not), and the outcome is the response of neoadjuvant chemotherapy NAC (effective versus not). We consider $1409$ radiomics features as potential confounders (see Web Appendix D).  We applied three confounder selection algorithms (SIS + OAL, SIS + GOAL and CBS) to evaluate the effect of the surgical stage on the treatment response of NAC. The first step of the algorithms use the SIS procedure of Tang et al (2022) to select the top $d_n =\lfloor n/log(n) \rfloor =22$ radiomics covariates (Fan et Lv, 2008). We then use the correlation thresholding (cutoff$=0.95$) to remove $10$ redundant variables (Becker et al. 2022).

Tables \ref{osres1} and \ref{osres2} present the results when the SIS + OAL, SIS + GOAL and CBS algorithms were used to fit the propensity score model to estimate the ATE of the surgical stage on the Effectiveness of NAC. We constructed $95\%$ normal confidence intervals for the ATE based on $10\, 000$ bootstrap resamples. In each bootstrap resample, we identified which radiomics features were selected for inclusion in the propensity score model based on the tolerance $10^{-8}$, which was used in Shortreed and Ertefaie (2017). That is, a radiomics feature is included in the PS model if the estimated coefficient is greater than $10^{-8}$ and excluded otherwise.

In Table \ref{osres1}, we report the point and normal confidence interval estimates. The ATE estimate for SIS + GOAL was $-0.307$ with a $95\%$ CI of $(-0.517, -0.097)$. The ATE estimate for SIS + OAL was $-0.317$ with a $95\%$ CI of $(-0.610, -0.023)$. The ATE estimate for CBS$^*$ was $-0.322$ with a $95\%$ CI of $(-2.361, 1.717)$. SIS + GOAL showed the shortest confidence interval compared to SIS+OAL and CBS. 

In Table \ref{osres2}, we present the proportion of times each radiomics feature was selected for inclusion in the PS model of SIS + GOAL, SIS + OAL or CBS. The range of the inclusion probabilities in the PS model of  SIS + GOAL was $58.1-98.7\%$, SIS + OAL was $50.9-98.5\%$ and CBS was $50.5-75.4\%$. While SIS + GOAL and SIS + OAL performed similarly for feature selection, SIS + GOAL showed slightly greater inclusion probability for each radiomics feature. The feature selection for CBS was very different to SIS + GOAL and SIS + OAL. The radiomics feature \textit{Large-Dependence Low-Gray-Level Emphasis.10} is known to be a strong predictor of NAC effectiveness (Zhang et al. 2021). It was selected with $99\%$ for both SIS + GOAL and SIS + OAL, while CBS rate was only $61\%$. 

\begin{table}[H]
\caption{SIS + OAL, SIS + GOAL and CBS estimators with correlation threshold $0.95$ and $B=10\,000$ for Osteosarcoma data. Note: CBS$^{*}$ is the CBS estimates between the 10th and 90th percentiles ($8\,000$ iterations) of the 10 000 bootstrap.}
\begin{center}
\begin{tabular}{lrrrrrrrr}
 \hline
  \hline
       & \textbf{ATE} & \textbf{Mean} & \textbf{Bias} & \textbf{SE} & \textbf{MSE} &\textbf{95\% CI} & \textbf{Length} \\ 
  \hline
  \hline
 \textbf{SIS + GOAL}   & -0.307 & -0.314 & \textbf{-0.007} & \textbf{0.107} & \textbf{0.011}  & \textbf{-0.517  to -0.097} & \textbf{0.419} \\
  \vspace{0.1cm}
  
  \textbf{SIS + OAL}    & -0.317 & -0.295 & 0.021  & 0.150 & 0.023  &-0.610  to -0.023 & 0.588  \\ 
  \vspace{0.1cm}
  
  \textbf{CBS$^{*}$}  & -0.322 & -0.537 & -0.215 & 1.041 &  1.129 & -2.361  to 1.717 & 4.079  \\ 
 \hline
\hline
\end{tabular}
\label{osres1}
\end{center}
\end{table}

\begin{table}[H]
\caption{Radiomic feature selection percentage (\%) for ATE using SIS + GOAL, SIS + OAL and CBS for $10\,000$ bootstrap resamples. The exposure is tumor stage (stage III versus non-stage III) and the outcome is effectiveness of treatment neoadjuvant chemotherapy (NAC).}

\begin{center}
\begin{tabular}{lrrrr}
\\
\hline
\hline
\textbf{Radiomic feature}  & \textbf{SIS + OAL} &  \textbf{CBS} & \textbf{SIS + GOAL}\\
\hline
\hline
Small-Dependence Low-Gray-Level Emphasis.7                                   &  80.7  &  68.5  &  83.7   \\
\vspace{0.1cm}

Sum Average.8                                                                &  81.2  &  67.4  &  85.6  \\
\vspace{0.1cm}

Zone Percentage.9                                                            &  84.9  &  55.7  &  87.5 \\
\vspace{0.1cm}

Cluster Tendency.10                                                          &  71.4  &  66.7  &  76.6 \\
\vspace{0.1cm}

Dependence Entropy.10                                                       &  83.9  &  51.5 &  87.2   \\
\vspace{0.1cm}

Large-Dependence Low-Gray-Level Emphasis.10                                 &  98.5  &  61.0 &  98.7   \\
\vspace{0.1cm}

Small-Dependence Low-Gray-Level Emphasis.10                                  &  95.0  &  50.5  &  96.3   \\
\vspace{0.1cm}

Short Run Low-Gray-Level Emphasis.10                                         &  93.4  &  62.2  &  94.6  \\
\vspace{0.1cm}

Long Run Low-Gray-Level Emphasis.10                                          &  96.4  &  68.6  &  97.3  \\
\vspace{0.1cm}

Variance.12                                                                  &  61.8  &  75.4  &  68.8   \\
\vspace{0.1cm}

Kurtosis.12                                                                  &  90.2  &  67.1  &  92.3  \\
\vspace{0.1cm}

Zone Entropy.12                                                              &  50.9  &  67.1  &  58.1 \\

\hline
\hline
\end{tabular}
\label{osres2}
\end{center}
\end{table}

\subsection{Gliosarcoma study\label{glioresults}}
In this gliosarcoma study, the exposure is Edema and the outcome is gliosarcoma (yes or no). We use 1303 radiomics features as potential confounders (see Web Appendix E). We examine the effect of the Edema on the gliosarcoma using the SIS + GOAL, SIS + OAL and CBS algorithms. In the first step, a number $d_n =\lfloor n/log(n) \rfloor =35$  radiomics features were selected. To remove the redundant features, we use the correlation thresholding function of Becker et al. (2022), using a cutoff of $0.85$. 

Tables \ref{gliores1} and \ref{gliores2} report the results when SIS + GOAL, SIS + OAL and CBS algorithms were used to fit the PS model to estimate the ATE of the Edema on the gliosarcoma. We performed $1\, 000$ bootstrap iterations to construct $95\%$ normal confidence intervals for the ATE. For each bootstrap iteration, we verified the set of radiomics features included in the PS model (with tolerance $10^{-8}$). 

In Table \ref{gliores1}, we present the point estimate and normal confidence interval for the ATE. The ATE estimate for SIS + GOAL was $0.242$ with a $95\%$ CI of $(0.069, 0.415)$. The ATE estimate for SIS + OAL was $0.227$ with a $95\%$ CI of $(0.042, 0.412)$. The ATE estimate for CBS$^*$ was $0.268$ with a $95\%$ CI of $(-0.118, 0.653)$. In this study also SIS + GOAL showed the shortest confidence interval compared to SIS + OAL and CBS. 


In Table \ref{gliores2}, we present the probability of inclusion of each radiomics feature in the PS model of  SIS + GOAL, SIS + OAL or CBS algorithms. In this study, the range of the selection rate for inclusion in the PS model of  SIS + GOAL was $65.4-99.9\%$, SIS + OAL was $54.5-99.8\%$ and CBS was $54.0-69.7\%$. We found that SIS + GOAL and SIS + OAL had similar performance for variable selection with slightly better rate for SIS + GOAL algorithm, while CBS performed differently.

\begin{table}[H]
\caption{SIS + OAL, SIS + GOAL and CBS estimator with correlation threshold $0.85$ and $B=1\, 000$ for Gliosarcoma data.  Note: CBS$^{*}$ is the CBS estimates between the 5th and 95th percentiles of the $1\, 000$ bootstrap.}
\begin{center}
\begin{tabular}{lrrrrrrrr}
 \hline
  \hline
       & \textbf{ATE} & \textbf{Mean} & \textbf{Bias} & \textbf{SE} & \textbf{MSE} &\textbf{95\%  CI} & \textbf{Length} \\ 
  \hline
  \hline
    \textbf{SIS + GOAL} & 0.242  &   0.240   & \textbf{-0.002}    & \textbf{0.088}   &  \textbf{0.008} &   \textbf{0.069   to  0.415}   &  \textbf{0.346}   \\
   \textbf{SIS + OAL} &  0.227  &   0.232  &   0.005   &  0.094  &  0.009     &  0.042  to  0.412  &  0.370   \\
  \textbf{CBS$^{*}$}  & 0.268   &  0.283   &  0.015   &  0.197  &  0.039 &  -0.118  to  0.653  &  0.770     \\
   \hline
\hline
\end{tabular}
\label{gliores1}
\end{center}
\end{table}

\begin{table}[H]
\caption{Radiomic feature selection percentage (\%) for ATE using SIS + GOAL, SIS + OAL and CBS  for $1\, 000$ bootstrap resamples. The exposure is Edema and the outcome is Gliosarcoma.}
\begin{center}
\begin{tabular}{lrrrr}
\\
\hline
\hline
\textbf{Radiomic feature}  & \textbf{SIS + OAL} &  \textbf{CBS} & \textbf{SIS + GOAL}\\
\hline
\hline
Original shape-MajorAxis                                                       & 58.2 &  57.9  &  65.4  \\
\vspace{0.1cm}

Original shape-Sphericity                                                      & 99.8  &  60.9  &  99.9  \\
\vspace{0.1cm}

Original glcm-JointEntropy                                                     & 98.2  &  56.8  &   98.7 \\
\vspace{0.1cm}

Original glcm-MaximumProbability                                               & 87.8  &  54.0  &   91.9 \\
\vspace{0.1cm}

LoG-sigma-3-0-mm-3D glszm-SizeZoneNonUniformity                                & 93.1  &  57.3  &   94.5 \\
\vspace{0.1cm}

LoG-sigma-5-0-mm-3D firstorder-RootMeanSquared                                 & 88.4  &  64.0  &   91.7 \\
\vspace{0.1cm}

Wavelet-LLL firstorder-Kurtosis                                                & 54.5 &  62.0  &   65.7 \\
\vspace{0.1cm}

Wavelet-LHH firstorder-Variance                                                & 77.1  &  61.9  &   82.0 \\
\vspace{0.1cm}

Wavelet-HHL firstorder-Skewness                                                & 68.1  &  58.8   &  77.1  \\
\vspace{0.1cm}

Wavelet-HHL gldm-LowGrayLevelEmphasis                                          & 71.2  & 69.7   &  78.1  \\
\vspace{0.1cm}

Squareroot glrlm-HighGrayLevelRunEmphasis                                      & 71.1  &  62.2  &   77.1 \\
\vspace{0.1cm} 

Squareroot glrlm-RunVariance                                                   & 94.0  &  62.3   &   95.5 \\

\hline
\hline
\end{tabular}
\label{gliores2}
\end{center}
\end{table}

\section{Discussion \label{dicuss}}
In this paper, we investigated the ability of ultra-high dimensional variable selection algorithms for causal inference to control confounding bias and improve statistical efficiency in radiomics data analysis. We first extended GOAL and OAL for ultra-high dimensional data analysis, SIS + GOAL and SIS + OAL. We then compare SIS + GOAL, SIS + OAL and CBS using simulation scenarios and two different radiomics datasets in cancer, osteosarcoma and gliosarcoma. Both simulation and radiomics data applications results showed that SIS + GOAL performed better than SIS + OAL and CBS. Our results showed that machine learning algorithms for causal inference are very useful for radiomics data analysis. In particular, machine learning algorithms methods which can achieve oracle property and collinearity simultaneously are more appropriate for radiomics data. Indeed, our findings offer new insights in the radiomics data analysis literature for both goal, causal inference and prediction modeling. 

As elucidated in Pearl and McKenzie (2018), there is a three-rung ladder for understanding the effects of exposures or interventions on outcomes.  While machine learning is popular with radiomics data, it constitutes step one of the ladder.  By contrast, the methods in this paper represent higher levels on the causal ladder of Pearl and McKenzie. 
\backmatter

\section*{Acknowledgements}
This work was funded by grants from  New Brunswick Innovation Foundation (NBIF) and the Grohne-Stepp Endowed Chair from the University of Colorado Cancer Center.
\section*{Data Availability Statement}
The data and software that support the findings of this paper are available in the Supporting Information section of this paper.

\section*{Supporting Information}
The {\tt{R}} code and  {\tt{R}} markdown documents to reproduce the two radiomics studies and the simulation study are available with this paper on Github:
\url{https://github.com/Ghoshlab/Radiomics_paper}.
\vspace*{-8pt}

\label{lastpage}

\end{document}